  \documentclass[12pt]{article}
  \usepackage{amssymb}
  \usepackage{latexsym}
  \usepackage{graphicx}

 \catcode`\@=11 \@addtoreset{equation}{section}\catcode`\@=12

 \def\barr{\left(\begin{array}}
 \def\earr{\end{array}\right)}

\newcommand{\R}{ {\mathbb R} }
\newcommand{\eps}{ \varepsilon }

\newcommand{\fnm}{\footnotemark}
\newcommand{\fnt}{\footnotetext}

\begin{document}

 \begin{center}

 \large \bf On exponential cosmological  type solutions
in  the   model with  Gauss-Bonnet  term and  variation of  gravitational constant

           \end{center}

 \vspace{0.3truecm}

 \begin{center}

 \normalsize\bf V. D. Ivashchuk \fnm[1]\fnt[1]{e-mail:
  ivashchuk@mail.ru} and A. A. Kobtsev \fnm[2]\fnt[2]{e-mail:
  speedyblackbird@gmail.com},

\vspace{0.3truecm}

 \it Center for Gravitation and Fundamental Metrology,
 VNIIMS, 46 Ozyornaya ul., Moscow 119361, Russia

 \it Institute of Gravitation and Cosmology,
 Peoples' Friendship University of Russia,
 6 Miklukho-Maklaya ul., Moscow 117198, Russia

\end{center}

 \begin{abstract}

  A $D$-dimensional  gravitational model with Gauss-Bonnet  term is
  considered. When  ansatz with diagonal  cosmological type metrics is adopted,
  we find solutions with exponential dependence of scale factors
  (with respect to ``synchronous-like'' variable)
  which  describe   an exponential expansion of ``our'' 3-dimensional factor-space
  and obey  the observational constraints on the temporal variation of effective gravitational constant $G$.
  Among them there are two exact solutions in dimensions $D = 22, 28$ with  constant $G$ and also
     an infinite series of solutions in dimensions $D \ge 2690$  with the
  variation of $G$ obeying the observational data.

 \end{abstract}

\clearpage

\section{Introduction}

Here we deal with $D$-dimensional gravitational model with the Gauss-Bonnet term.
The action reads
\begin{equation}
 S =  \int_{M} d^{D}z \sqrt{|g|} \{ \alpha_1 R[g] +
             \alpha_2  {\cal L}_2[g] \},
   \label{1.1}
 \end{equation}
where $g = g_{MN} dz^{M} \otimes dz^{N}$ is the metric defined on
the manifold $M$, ${\dim M} = D$, $|g| = |\det (g_{MN})|$ and

\begin{equation}
  {\cal L}_2 = R_{MNPQ} R^{MNPQ} - 4 R_{MN} R^{MN} +R^2
   \label{1.2}
 \end{equation}
is the standard Gauss-Bonnet term. Here $\alpha_1$ and $\alpha_2$
are non-zero constants.

 Earlier the appearance of the  Gauss-Bonnet
term  was motivated by string theory
 \cite{Zwiebach,GBstrings1,GBstrings2,GBstrings3,GBstrings4}.
 At present, the (so-called) Einstein-Gauss-Bonnet (EGB) gravitational model and
 its modifications are intensively used in  cosmology,
 see \cite{NojOd0,CElOdZ} (for $D =4$),
 \cite{Ishihara,Deruelle,ElMakObOsFil,BambaGuoOhta,TT,KirMPTop,PTop,KirMak,ChPavTop}
 and references therein,  e.g. for explanation  of  accelerating
 expansion of the Universe following from
 supernovae (type Ia) observational data \cite{Riess,Perl,Kowalski}.
 Certain exact solutions in multidimesional EGB cosmology
 were obtained in  \cite{Ishihara}-\cite{ChPavTop}  and
 some other papers.

  Here we are dealing with the cosmological type solutions with diagonal
 metrics (of Bianchi-I-like type) governed by $n$ scale factors
 depending upon one variable, where $n > 3$. Moreover, we restrict ourselves by the
 solutions with exponential dependence of scale factors
  (with respect to ``synchronous-like'' variable $\tau$)

  \begin{equation}
  \label{ah}
  a_i(\tau) \sim \exp{(h^i \tau )},
  \end{equation}
   $i = 1, \dots, n$; $D = n+1$,
  with the aim to find solutions
  describing   an exponential isotropic expansion of  3-dimensional flat factor-space, i.e.
  with
  \begin{equation}
  \label{HHH}
  h^1 = h^2 =h^3 = H >0,
  \end{equation}
   and small enough variation of the effective
  gravitational constant $G$, which is proportional to the inverse volume scale factor
   of the internal space, i.e.

  \begin{equation}
  \label{G0}
    G \sim \prod_{i=4}^{n} [a_i(\tau)]^{-1} \sim \exp{(- Int \ \tau) },
   \end{equation}
where here and in what follows we denote
 \begin{equation}
  \label{Int}
   Int = \sum_{i=4}^{n} h^i .
   \end{equation}
 see  \cite{BIM,AIKM,IKMN,Mel,IvMel-14} and references therein.
 We call $h^i= \dot{a_i}/ a_i$  as ``Hubble-like'' parameter
 corresponding to  $i$-th subspace.

In  cosmological model under consideration with anisotropic ``internal space'',
we get for dimensionless parameter of temporal variation of $G$
the following relation from (\ref{G0}) and (\ref{Int})
\begin{equation}
\frac{\dot{G}}{GH} =  -Int / H,
\label{G}
\end{equation}
where $H$ is the Hubble parameter.

As for the experimental data, the variation of the gravitational constant is allowed
at the level of $10^{-13}$ per year and less.
We use the following constraint on the magnitude of the dimensionless variation of the
 gravitational constant

 \begin{equation}
 \label{G1}
  - 0,65 \cdot 10^{-3} < \frac{\dot{G}}{GH} < 1,12 \cdot 10^{-3},
 \end{equation}
which comes from the most stringent limitation
on $G$-dot obtained by the set of ephemerides \cite{Pitjeva}

\begin{equation}
 \label{G2}
\dot{G}/G = (0.16 \pm 0.6) \cdot 10^{-13} \ year^{-1}
\end{equation}
allowed at 95\% confidence (2-$\sigma$) level
and the present value of the Hubble parameter \cite{Ade}
(which characterizes the rate of expansion of the observable Universe)
 \begin{equation}
 \label{H}
  H_0 =  (67,80 \pm 1,54) \ km/s \ Mpc^{-1} =  (6.929  \pm 0,157) \cdot 10^{-11} \ year^{-1},
 \end{equation}
  with 95\% confidence level. It shold be noted that the original result for $H_0$ in \cite{Ade}
  (for the Planck best-fit cosmology including external data set) was presented
  at 68\% confidence (1-$\sigma$) level.  In restriction (\ref{G1}) we use the lower allowed
  value for  $H_0$ in (\ref{H}) in order to obtain the confidence  level more than 95\% .

Thus, we are seeking here the cosmological solutions which obey the relations
(\ref{ah})-(\ref{G1}), listed above.

  The paper is organized as follows. In Section 2
  the equations of motion for  $D$-dimensional EGB model are considered.
  For  diagonal  cosmological type metrics
  the equations of motion are equivalent to a set of Lagrange equations
  corresponding to certain ``effective'' Lagrangian \cite{Iv-09,Iv-10}
  (see also \cite{Deruelle,Pavl}).
 In Section 3  some   cosmological solutions with exponential behavior of scale factors satisfying
the restriction (\ref{G1})   are obtained for
two isotropic factor spaces and positive  value of  $\alpha = \alpha_2/\alpha_1$.

\section{The cosmological type model and its effective Lagrangian}

 \subsection{The set-up }

 Here we consider the manifold
 \begin{equation}
   M = \R_{*}  \times  M_1 \times ... \times M_n , \label{2.1}
 \end{equation}
  with the metric
 \begin{equation}
  g= w e^{2{\gamma}(u)} du \otimes du  +
 \sum_{i=1}^{n} e^{2\beta^i(u)} \eps_{i} dy^i \otimes dy^i,
 \label{2.2}
 \end{equation}
 where $w = \pm 1$,   $ \eps_{i}= \pm 1$, $i = 1, \dots, n$,
 and   $M_1, ...,  M_n$  are one dimensional manifolds (either $\R$ or $S^1$).
 Here and in what follows
  $\R_{*} = (u_{-},u_{+})$ is an open subset in $\R$.
 The functions ${\gamma}(u)$ and
 $\beta^i(u)$,  $i = 1,\ldots, n$, are smooth on
 $\R_{*} = (u_{-},u_{+})$.

 For $w =  -1$, $ \eps_{1}=  ... = \eps_{n} = 1$
 the metric (\ref{2.2}) is a cosmological one, while
 for $w =  1$, $ \eps_{1}= -1$,
 $\eps_{2} = ... =\eps_{n} = 1$ it describes certain static
 configurations.

 For physical applications  we put $M_1 =M_2 = M_3 = \R$, while
   $M_4, ..., M_n$ will be considered to be compact ones (i.e. coinciding with
 $S^1$).

 The integrand in  (\ref{1.1}), when the
 metric (\ref{2.2}) is substituted, reads as follows
   \begin{equation}
    \sqrt{|g|} \{ \alpha_1 R[g] +
               \alpha_2  {\cal L}_2[g] \} = L + \frac{df}{du},
    \label{2.3}
   \end{equation}
 where
    \begin{eqnarray}
    L = \alpha_1 L_1 +  \alpha_2 L_2,
    \label{2.4}\\
      L_1 = (-w) e^{-\gamma + \gamma_0} G_{ij} \dot{\beta}^i
      \dot{\beta}^j,
         \label{2.5}   \\
       L_2  = - \frac{1}{3}  e^{- 3 \gamma + \gamma_0}
         G_{ijkl} \dot{\beta}^i \dot{\beta}^j \dot{\beta}^k
         \dot{\beta}^l,
              \label{2.6}  \\
       \gamma_0 = \sum_{i =1}^{n} \beta^i
    \end{eqnarray}
   and
  \begin{eqnarray}
       G_{ij} = \delta_{ij} -1,
         \label{2.10}   \\
       G_{ijkl}  = (\delta_{ij} -1)(\delta_{ik} -1)(\delta_{il} -1)
       (\delta_{jk} -1)(\delta_{jl} -1)(\delta_{kl} -1)
       \label{2.11}
      \end{eqnarray}
      are respectively the components of two  metrics on
      $\R^{n}$ \cite{Iv-09,Iv-10}.   The first one is the well-known ``minisupermetric'' - 2-metric
      of  pseudo-Euclidean signature: $<v_1,v_2> = G_{ij}v^i_1 v^j_2$
      and the second one is the Finslerian 4-metric:
      $<v_1,v_2,v_3,v_4> = G_{ijkl}v^i_1 v^j_2 v^k_3 v^l_4$,
      $v_s = (v^i_s) \in \R^n$,
      where $<.,.>$ and $<.,.,.,.>$ are respectively
      $2$- and $4$-linear symmetric forms on $\R^n$.
     Here we denote $\dot{A} = dA/du$ etc.
     The function $f(u)$ in (\ref{2.3}) is irrelevant
     for our consideration (see \cite{Iv-09,Iv-10}).

      The derivation of (\ref{2.4})-(\ref{2.6}) is based on
        the following identities \cite{Iv-09,Iv-10}:
      \begin{eqnarray}
       G_{ij}v^i v^j = \sum_{i =1}^{n} (v^i)^2 -
        (\sum_{i =1}^{n} v^i )^2,
         \label{2.12}   \\
       G_{ijkl}v^i v^j v^k v^l  = (\sum_{i =1}^{n} v^i )^4
        - 6 (\sum_{i =1}^{n} v^i )^2
        \sum_{j =1}^{n} (v^j)^2
         \nonumber \\
        +     3 ( \sum_{i =1}^{n} (v^i)^2 )^2
        + 8  (\sum_{i =1}^{n} v^i )
        \sum_{j =1}^{n} (v^j)^3  - 6 \sum_{i =1}^{n} (v^i)^4.
       \label{2.13}
      \end{eqnarray}

      It follows immediately from the definitions
      (\ref{2.10}) and (\ref{2.11}) that
      \begin{eqnarray}
       G_{ij}v^i v^j = -2 \sum_{i < j} v^i v^j,
         \label{2.14}   \\
       G_{ijkl}v^i v^j v^k v^l  = 24 \sum_{i < j < k < l} v^i v^j
       v^k v^l.
       \label{2.15}
      \end{eqnarray}

  \subsection{The equations of motion }

    The equations of motion corresponding to the action (\ref{1.1})
  have the following form
  \begin{equation}
   {\cal E}_{MN} = \alpha_1 {\cal E}^{(1)}_{MN}
     + \alpha_2 {\cal E}^{(2)}_{MN} = 0,
   \label{1.3e}
 \end{equation}
  where
  \begin{eqnarray}
   {\cal E}^{(1)}_{MN} = R_{MN} - \frac{1}{2} R g_{MN},
   \label{1.3a} \\
   {\cal E}^{(2)}_{MN} = 2(R_{MPQS}R_N^{\ \ PQS} - 2 R_{MP} R_N^{\ \ P}
   \nonumber \\
   -2 R_{MPNQ} R^{PQ} + R R_{MN}) -  \frac{1}{2} {\cal L}_2  g_{MN}.
   \label{1.3b}
   \end{eqnarray}

    It was  shown in  \cite{Iv-10} that the field equations (\ref{1.3e}) for the metric
    (\ref{2.2}) are equivalent to the Lagrange equations
    corresponding to the Lagrangian $L$ from (\ref{2.4}).

    Thus, equations (\ref{1.3e}) read as follows
    \begin{eqnarray}
       w \alpha_1  G_{ij} \dot{\beta}^i \dot{\beta}^j
        + \alpha_2  e^{- 2 \gamma}
         G_{ijkl} \dot{\beta}^i \dot{\beta}^j \dot{\beta}^k
         \dot{\beta}^l = 0,  \label{2.17} \\
         \frac{d}{du}[ - 2w \alpha_1  G_{ij} e^{-\gamma +
         \gamma_0}
          \dot{\beta}^j
        -  \frac{4}{3} \alpha_2 e^{- 3 \gamma + \gamma_0}
         G_{ijkl}  \dot{\beta}^j \dot{\beta}^k
         \dot{\beta}^l] - L = 0,   \label{2.18}
      \end{eqnarray}

     $i = 1,\ldots, n$. Due to (\ref{2.17})
               $L=  -w \frac{2}{3} e^{-\gamma + \gamma_0}
      \alpha_1  G_{ij} \dot{\beta}^i  \dot{\beta}^j$.

\subsection{Reduction to an autonomous system of first order differential equations}

   Now we put $\gamma = 0$ and   denote $u = \tau$, where $\tau$ is a ``synchronous-like'' variable.
 By introducing ``Hubble-like'' variables $h^i = \dot{\beta}^i$,
   the eqs. (\ref{2.17}) and  (\ref{2.18}) may be rewritten as  follows

  \begin{eqnarray}
       w \alpha_1  G_{ij} h^i h^j
        + \alpha_2   G_{ijkl} h^i h^j h^k h^l = 0,  \label{5.1} \\
          \left[ -2 w  \alpha_1  G_{ij} h^j
        -  \frac{4}{3} \alpha_2  G_{ijkl}  h^j h^k h^l \right] \sum_{i=1}^nh^i
        \qquad \nonumber \\
          + \frac{d}{d\tau} \left[ -2 w  \alpha_1  G_{ij} h^j
           -  \frac{4}{3} \alpha_2  G_{ijkl}  h^j h^k h^l \right]
           - L    = 0,   \label{5.2}
      \end{eqnarray}
     $i = 1,\ldots, n$, where
      \begin{equation}
       L =  -w \alpha_1 G_{ij} h^i h^j
                - \frac{1}{3} \alpha_2   G_{ijkl} h^i h^j h^k h^l.
         \label{5.1a}
       \end{equation}
     Due to (\ref{5.1})
       $L = -  \frac{2}{3} w \alpha_1  G_{ij} h^i h^j$.

      Thus,  we are led to the   autonomous system of the first-order differential equations on
        $h^1(\tau), ..., h^n(\tau)$  \cite{Iv-10,Iv-10}.

     Here and in what follows  we  use relations (\ref{2.12}), (\ref{2.13})
     and the following formulas
     \begin{eqnarray}
       G_{ij}v^j = v^i - S_1,
         \label{5.3}   \\
       G_{ijkl} v^j v^k v^l
       = S_1^3  + 2 S_3 -3 S_1 S_2  +  3 (S_2  - S_1^2)  v^i
         +  6 S_1 (v^i)^2 - 6(v^i)^3,
       \label{5.4}
      \end{eqnarray}
     $i = 1,\ldots, n$, where $S_k = S_k (v) = \sum_{i =1}^n
     (v^i)^k$.

\subsection{Solutions with constant  $h^i$  }

      In this paper we deal with the following solutions to
     equations  (\ref{5.1}) and (\ref{5.2})

       \begin{equation}
             h^i (\tau) = v^i,  \label{5.4v}
       \end{equation}
       with constant $v^i$,   which correspond to the solutions
      \begin{equation}
      \beta^i = v^i \tau +
      \beta^i_0,  \label{5.4a}
      \end{equation}
       where $\beta^i_0$ are constants, $i = 1,\ldots, n$.

      In this case we obtain the metric (\ref{2.2})
      with the exponential dependence of scale  factors
      \begin{equation}
        g= w d \tau \otimes d \tau  +
        \sum_{i=1}^{n} \eps_{i} B_i^2 e^{2v^i \tau} dy^i \otimes dy^i,
        \label{5.4m}
        \end{equation}
       where $w = \pm 1$, $ \eps_i= \pm 1$ and
        $B_i > 0$ are arbitrary constants.

      For the fixed point  $v = (v^i)$ we have the set of  polynomial equations
          \begin{eqnarray}
         G_{ij} v^i v^j
         - \alpha   G_{ijkl} v^i v^j v^k v^l = 0,  \label{5.5} \\
          \left[ 2   G_{ij} v^j
        -  \frac{4}{3} \alpha  G_{ijkl}  v^j v^k v^l \right] \sum_{i=1}^n v^i
        -  \frac{2}{3}   G_{ij} v^i v^j  = 0,   \label{5.6}
      \end{eqnarray}
     $i = 1,\ldots, n$, where  $\alpha = \alpha_2(-w)/\alpha_1$.
      For $n > 3$ this is a set of forth-order polynomial
      equations.

      The trivial solution $v = (v^i) = (0, ..., 0)$ corresponds
      to  a flat metric $g$.

      For any non-trivial solution $v$ we have
      $\sum_{i=1}^n v^i  \neq 0$ (otherwise one gets
      from (\ref{5.6}) $G_{ij} v^i v^j = \sum_{i =1}^{n} (v^i)^2 -
        (\sum_{i =1}^{n} v^i)^2 = 0$ and hence $v = (0, \dots, 0)$).

      The set of equations (\ref{5.5}) and (\ref{5.6})
      has an isotropic solution $v^1 = ... = v^n = H$,
      where
       \begin{equation}
        n(n -1)H^2 + \alpha n(n -1)(n -2)(n -3) H^4 = 0.
         \label{5.7}
        \end{equation}
         For $n = 1$:  $H$ is arbitrary and for $n = 2,3$: $H =0$.

         When $n > 3$,  the non-zero solution
         to eq. (\ref{5.7}) exists only if $\alpha  < 0$ and
         in this case \cite{Iv-09,Iv-10}
         \begin{equation}
         H = \pm \frac{1}{\sqrt{|\alpha| (n -2)(n -3)}}.
         \label{5.8}
         \end{equation}

 In cosmological case ($w  = -1$) this
 solution takes place when $\alpha_2/\alpha_1 = \alpha< 0$.

The isotropic solution for $n > 3$ gives rise
to a very large value of $\dot{G}/G  = (n-3) H$, which is forbidden by observational restrictions.

  It was shown in \cite{Iv-09,Iv-10} that there are no more than
 three different  numbers among  $v^1,...,v^n$. \fnm[2]\fnt[2]{ At the moment
we were unable to find  solutions with three different real ``Hubble-like'' parameters.}

\section{Examples of cosmological  solutions obeying the restriction on the variation of $G$}

 In this section we  consider some
 solutions to the set of equations (\ref{5.5}), (\ref{5.6}) of the following form
 $v =(H, \ldots, H, h, \ldots, h)$,
  where $H$  the ``Hubble-like'' parameter corresponding  to $m$-dimensional isotropic subspace
with $m > 3$ and $h$ is the ``Hubble-like'' parameter   corresponding to $l$-dimensional isotropic subspace, $l>2$.

 These solutions should satisfy  the following conditions:
\begin{enumerate}
\item mandatory:

\begin{enumerate}
 \item $H$ and $h$ are real numbers,
 \item $H > 0$, $h < 0$;

 \end{enumerate}

\item desirable:

\begin{enumerate}

 \item $Int = (m-3)H+lh<0$;

 \item $- 0,65 \cdot 10^{-3} < \frac{\dot{G}}{GH}= -((m-3)+\frac{lh}{H}) < 1,12 \cdot 10^{-3}$.

\end{enumerate}

\end{enumerate}

The first inequality $H > 0$ in the mandatory condition is necessary for  a description of accelerated expansion of 3-dimensional subspace, which may describe our Universe, while the second inequality $h < 0$ excludes an enormous (of Hubble's order) variation  $\dot{G}/G$ for  $h \ge 0$ and $m > 3$.

The first desirable condition means that the volume scale factor of the internal space
 $V(\tau)= B \exp(((m-3)H+lh) \tau )$, where $B >0$ is constant, decreases over time.
This condition is a sort of weak extension of a possible restriction for $m=3$ coming from the unobservability of the ``internal space'' for all $\tau > \tau_0$. It is also desirable since the negative value  of the
parameter  $Int$ is more probable due to more probable positive value of $\dot{G}/G = -Int$, see
 (\ref{G2}).

The second desirable condition may be also  rewritten  by using
the parameter $Var = |\frac{\dot{G}}{GH}| = |(m-3)+\frac{lh}{H}| $:
\begin{eqnarray}
Var  < 1,12 \cdot 10^{-3}, \quad  {\rm for} \quad  Int \le 0; \label{Var1} \\
Var <  0,65 \cdot 10^{-3}, \quad {\rm for} \quad Int \ge 0. \label{Var2}
\end{eqnarray}

Here we consider the simplest case, when internal spaces (apart the expansion factors) are flat. The consideration of curved internal spaces will drastically change the equations of motion and may break the existence of solutions with exponential dependence of scale factors. Anyway the inclusion into consideration of  curved internal spaces may be worth but need a special treatment, which may be done   in a separate work.

\subsection{The dependence of ``Hubble-like'' parameters on $m$ and $l$}

The total dimension of the considered space is $D = n+1=(m+l)+1$, where we have $m$ dimensions expanding with the Hubble parameter $H >0$ and $l$ dimensions contracting  with the ``Hubble-like''  parameter $h <0$.

According to this, we rewrite the set of polynomial equations (\ref{5.5}), (\ref{5.6}), using the formulas (\ref{5.3}), (\ref{5.4}), as follows:
\begin{eqnarray}
H^{2}(m-m^{2})+h^{2}(l-l^{2})-2mlHh \nonumber \\
-\alpha(H^{4}m(m-1)(m-2)(m-3)+h^{4}l(l-1)(l-2)(l-3) \nonumber \\
+4H^{3}hm(m-1)(m-2)l+4h^{3}Hl(l-1)(l-2)m \nonumber \\
+6H^{2}h^{2}m(m-1)l(l-1))=0, \label{U0}
\end{eqnarray}
\begin{eqnarray}
m(1-m)H^{2}-(1/2)lh^{2}(1+2l)+2lHh((3/4)-m) \nonumber \\
-\alpha(H^{4}m(m-1)(m-2)(m-3)+H^{3}hl(m-1)(m-2)(4m-3) \nonumber \\
+3H^{2}h^{2}l(m-1)(2lm-2l-m) \nonumber \\
+Hh^{3}l(l-1)(4lm-3l-2m)+h^{4}l^{2}(l-1)(l-2))=0, \label{UH}
\end{eqnarray}
\begin{eqnarray}
l(1-l)h^{2}-(1/2)mH^{2}(1+2m)+2mHh((3/4)-l)\nonumber\\
-\alpha(h^{4}l(l-1)(l-2)(l-3)+h^{3}H m(l-1)(l-2)(4l-3)\nonumber\\
+3h^{2}H^{2} m(l-1)(2lm-2m-l)\nonumber\\+h H^{3} m(m-1)(4lm-3m-2l)+H^{4}m^{2}(m-1)(m-2))=0.
\label{Uh}
\end{eqnarray}

Here we put  for simplicity $\alpha = \pm 1$ but keep in mind that general $\alpha$-dependent
solution has the following form

\begin{equation}
\label{Hh}
H(\alpha) = H |\alpha|^{-1/2}, \qquad  h(\alpha) = h |\alpha|^{-1/2}.
\end{equation}

Due to these relations the parameter $\dot{G}/(GH)$ does not depend upon
$|\alpha|$ and hence our simplification is a reasonable one. For any solution $(H,h)$ with
$\alpha = \pm 1$ we can find a proper $\alpha$ which will be in agreement with
the present value of the  Hubble parameter $H_0$ (see (\ref{H}))
 \begin{equation}
  \label{H0}
  H |\alpha|^{-1/2} = H_0.
 \end{equation}

Our numerical analysis (based on Maplesoft Maple) show us that (generically) there are 11 solutions of these equations
(for $m > 3$ and $l \ge 3$).

{\bf I)} The first to mention is, obviously, the zero solution $H_1=h_1=0$.

{\bf II)} Two other solutions are isotropic ones:
\begin{enumerate}
\item if $\alpha=1$, then $H= h =  \pm  \sqrt{\frac{1}{(l^2+2lm+m^2-5l-5m+6)}} \cdot i$.
We are led to pure imaginary  isotropic solutions obeying (\ref{5.7}).
For example, when $m=9$ and $l=6$ we receive
\begin{enumerate}
\item $H_2= h_2 =\frac{1}{2}\sqrt{\frac{1}{39}}\cdot i$;
\item $H_3=h_3=-\frac{1}{2}\sqrt{\frac{1}{39}}\cdot i$;
\end{enumerate}
\item if $\alpha=-1$, then $H= h = \pm \sqrt{\frac{1}{(l^2+2lm+m^2-5l-5m+6)}}$.
We are led to  isotropic solutions (\ref{5.8}).
When $m=9$ and $l=6$ the solutions are:
\begin{enumerate}
\item $H_2=h_2=\frac{1}{2}\sqrt{\frac{1}{39}}$;
\item $H_3=h_3=-\frac{1}{2}\sqrt{\frac{1}{39}}$.
\end{enumerate}
\end{enumerate}

{\bf III)} For $\alpha=\pm1$ the remaining eight solutions are  roots of the following
two equations which are given by Maple:
\begin{eqnarray}
P(H;m,l) =  64(m-2)(m-1)^2(m-2+l)(l^2m+lm^2-2l^2
\nonumber\\+2lm-2m^2)(-3+l+m)^2\cdot H^8 \nonumber\\
\mp128(m-1)^2(-3+l+m)(l^2m(l+m)^2-2l(l+m)^3
\nonumber \\
+(l+m)(8l^2+lm+2m^2)  -10l^2+4lm-4m^2) \cdot H^6
\nonumber\\
+16(m-1)(5l^5m+10l^4m^2+5l^3m^3-6l^5-38l^4m
\nonumber \\-49l^3m^2-17l^2m^3+32l^4
\nonumber \\
 +111l^3m+75l^2m^2+14lm^3-70l^3-130l^2m-14lm^2
 \nonumber\\-8m^3+68l^2+4lm+8m^2)\cdot H^4
 \nonumber\\
 \mp 16l(m-1)(l^4+l^3m-3l^3-5l^2m+5l^2+8lm-7l-2m)\cdot H^2
\nonumber\\ +l^5-3l^3-2l^2=0, \label{EqqPos}
\end{eqnarray}
\begin{eqnarray}
P(h;l,m) = 0.
\label{EqqPos1}
\end{eqnarray}

The second equation is obtained from the first one just by  swapping  parameters $m$ and $l$ and replacing $H$ by $h$.
The solutions to eqs. (\ref{EqqPos}), (\ref{EqqPos1})  should be  substituted
into eqs. (\ref{U0}), (\ref{UH}), (\ref{Uh}), in order to find  the solutions  $(H,h)$ under consideration.

The closed-form expression for the solution  in general case (for any $m$ and $l$) seems to be  very bulky.
So, we use Maplesoft Maple to find  solutions for certain $m$ and $l$ and test  some general  features of these solutions\fnm[4]\fnt[4]{Here
we are led to some common features of the solutions just by numerical calculations for a restricted range of numbers  $m$ and $l$.}:

\begin{enumerate}
\item for $\alpha=1$ in common case we have two pairs of real and quad of complex solutions which differ in  signs. (See footnote 4.)
 For $m=9$ and $l=6$ we receive:
\begin{enumerate}
\item $H_4\approx-0.2597$ and $h_4\approx0.2826$;
\item $H_5\approx0.2597$ and $h_5\approx-0.2826$;
\item $H_6\approx-0.1610$ and $h_6\approx0.4004$;
\item $H_7\approx0.1610$ and $h_7\approx-0.4004$;
\item $H_8\approx-0.0913-0.0464\cdot i$ and $h_8\approx0.1456-0.0449\cdot i$;
\item $H_9\approx-0.0913+0.0464\cdot i$ and $h_9\approx0.1456+0.0449\cdot i$;
\item $H_{10}\approx0.0913-0.0464\cdot i$ and $h_{10}\approx-0.1456-0.0449\cdot i$;
\item $H_{11}\approx0.0913+0.0464\cdot i$ and $h_{11}\approx-0.1456+0.0449\cdot i$;
\end{enumerate}
\item  for $\alpha=-1$ ($m > 3$ and $l \ge 3$) there are no (extra) real solutions. (See footnote 4.)

 In the case of $m=9$ and $l=6$ we  obtain:
\begin{enumerate}
\item $H_4\approx-0.2597\cdot i$ and $h_4\approx0.2826\cdot i$;
\item $H_5\approx0.2597\cdot i$ and $h_5\approx-0.2826\cdot i$;
\item $H_6\approx0.1610\cdot i$ and $h_6\approx-0.4004\cdot i$;
\item $H_7\approx-0.1610\cdot i$ and $h_7\approx0.4004\cdot i$;
\item $H_8\approx-0.0464-0.0913\cdot i$ and $h_8\approx-0.0449+0.1456\cdot i$;
\item $H_9\approx-0.0464+0.0913\cdot i$ and $h_9\approx-0.0449-0.1456\cdot i$;
\item $H_{10}\approx0.0464-0.0913\cdot i$ and $h_{10}\approx0.0449+0.1456\cdot i$;
\item $H_{11}\approx0.0464+0.0913\cdot i$ and $h_{11}\approx0.0449-0.1456\cdot i$.
\end{enumerate}
\end{enumerate}

\vfill

It can be  seen that none of the solutions for $\alpha=-1$ satisfies our mandatory conditions written in the beginning of this section.

As for $\alpha=1$, the solutions III.1.b and III.1.d are real and $H > 0$,   $h < 0$.
It can be verified that in these cases $Int= (m-3)H+lh < 0$.

Now we have  to calculate the  variation of the gravitational constant. For $m=9$ and $l=6$ we get:
\vfill

\begin{math} Var_5= |\frac{\dot{G}}{GH}|_5= |{(m-3)+\frac{l\cdot h_5}{H_5}}| \approx 0.535826;
\end{math}

\vfill

\begin{math} Var_7= |\frac{\dot{G}}{GH}|_7= |{(m-3)+\frac{l\cdot h_7}{H_7}}| \approx 8.914741.
\end{math}
\vfill

The first variation is lower: $Var_5 < Var_7$,  for $m=9$ and $l=6$.
This inequality  seems to take place  for any  $m > 3$ and $l \ge 3$. At the moment  a rigorous proof of this fact
is absent while certain  numerical calculations support it. Anyway, here  we will  focus on
 the solution $(H_5, h_5)$,  which we consider as a more interesting (for our applications)  then $(H_7, h_7)$.
Further we will write $H$ and $h$ instead of $H_5$ and $h_5$ in common case.

 We can plot behaviour of  parameters $H$, $h$, $Int$ and $Var$, for example, keeping fixed $m=8$ and $m=10$ and raising $l$ from 5 to 100 by 5. See Figure 1 and Figure 2.

\begin{figure}[h]
\begin{minipage}[h]{0.49\linewidth}
\center{\includegraphics[width=1\linewidth]{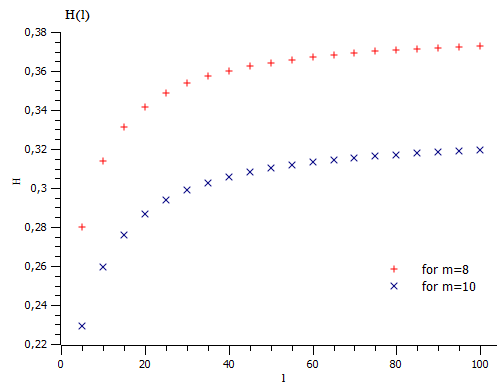}}
\end{minipage}
\hfill
\begin{minipage}[h]{0.49\linewidth}
\center{\includegraphics[width=1\linewidth]{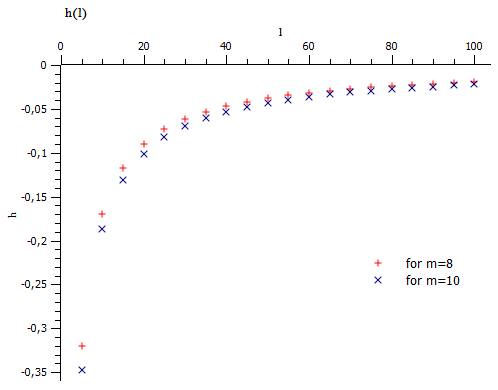}}
\end{minipage}
\caption{Behavior of  ``Hubble-like'' parameters $H$ and $h$ for fixed $m=8$ and $m=10$ while  $l$
is changing.}
\label{ris:image1}
\end{figure}

\begin{figure}[h]
\begin{minipage}[h]{0.49\linewidth}
\center{\includegraphics[width=1\linewidth]{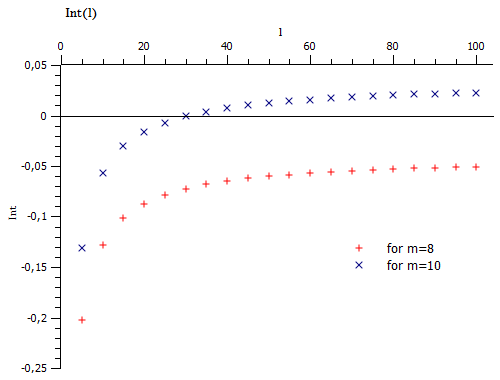}}
\end{minipage}
\hfill
\begin{minipage}[h]{0.49\linewidth}
\center{\includegraphics[width=1\linewidth]{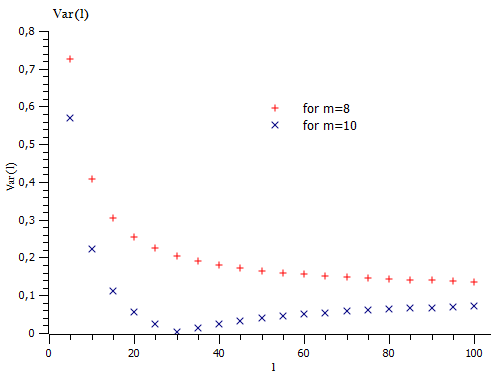}}
\end{minipage}
\caption{Behavior of the internal space parameter $Int$ and the variation of $G$ parameter  $Var$ for fixed $m=8$ and $m=10$ while  $l$ is changing.}
\label{ris:image2}
\end{figure}

\subsection{The limiting values of  $H$,  $hl$, $Int$ and $Var$   for fixed $m \le 9$ }
\vfill
When $m \le 9$ the internal space parameter $Int$ remains negative, that means that the first desirable condition is satisfied for any $l$.
The variation of $G$ parameter is monotonically decreasing with the increase of $l$. Moreover, we get
finite limits for $H$ and $h l$ as $l \to + \infty$.
In this subsection we obtain these and other limits (for $Int$ and $Var$)
   for fixed $m \le 9$.

   Now let us rewrite (\ref{EqqPos}) and (\ref{EqqPos1}) for $H$ and $hl$ keeping only the terms with higher degrees of $l$:

\vfill
\begin{eqnarray}
64(m-1)^2(m-2)^2l^5\cdot H^8-128(m-2)(m-1)^2l^5\cdot H^6\nonumber\\+16(m-1)(5m-6)l^5\cdot H^4-16(m-1)l^5\cdot H^2+l^5=0,
\label{SolH}
\end{eqnarray}
\begin{eqnarray}
64(m-1)(l\cdot h)^8-128(m^2-2m+2)(l\cdot h)^6\nonumber\\+(80m^3-272m^2+224m-128)(l\cdot h)^4
\nonumber\\-16m(m^3-5m^2+8m-2)(l\cdot h)^2 \nonumber \\+m^5-3m^3-2m^2=0.
\label{Solh}
\end{eqnarray}
\vfill
Solving these equations we  find the limiting values:
\vfill
\begin{eqnarray}
\lim_{l \to \infty}{H}=\frac{1}{2}\sqrt{\frac{2m-2+\sqrt{2m^2-2m}}{m^2-3m+2}},
\label{LimH}
\end{eqnarray}
\begin{eqnarray}
\lim_{l \to \infty}{h\cdot l}=-\frac{1}{2}\Bigr(\frac{1}{m-2}(m\sqrt{2m(m-1)}+2m^2
 \nonumber \\
-2\sqrt{\frac{2((m+2)\sqrt{m(m-1)}+m^2\sqrt{2})(m-1)^2}{\sqrt{m(m-1)}}} \nonumber \\
-4m+4)\Bigl)^{\frac{1}{2}},
\label{Limhl}
\end{eqnarray}
\begin{eqnarray}
\lim_{l \to \infty}{Var}= \Bigr|(m-3)+\frac{\lim_{l \to \infty}{h\cdot l}}{\lim_{l \to \infty}{H}} \Bigl|.
\label{LimhVar}
\end{eqnarray}
\vfill

We   fill  the Table 1 by calculated values for $3 \le m \le 9$:
\clearpage
\begin{table}[ht]
\begin{tabular}{|c|p{0.2\textwidth}|p{0.2\textwidth}|
p{0.2\textwidth}|p{0.2\textwidth}|}
\hline m & $\lim_{l\to\infty}{H}$ & $\lim_{l\to\infty}{l\cdot h}$ & $\lim_{l\to\infty}{Int}$ & $\lim_{l\to\infty}{Var}$ \\ \hline
3 & 0.9659258265 & -0.7630807575
& -0.7630807575 & 0.7899993318 \\ \hline
4 & 0.6738873385 & -1.093021916 & -0.4191345775
& 0.621965355 \\ \hline
5 & 0.5462858555 & -1.352249104 & -0.259677393
& 0.475350754 \\ \hline
6 & 0.4709825726 & -1.574449592& -0.161501874
& 0.342904141 \\ \hline
7 & 0.4199717390 & -1.772664074 & -0.092777118
& 0.220912766 \\ \hline
8 & 0.3825276619 & -1.953613607 & -0.040975297
& 0.107117214 \\ \hline
9 & 0.3535533906 & -2.121320344 & 0
& 0 \\ \hline
\end{tabular}
\caption{The limiting values of $H$, $hl$, $Int$ and $Var$ parameters as $l \to \infty$.}
\label{sample_table}
\end{table}

For $3 \le m \le 8$ the limiting values of  $Var$-parameter are too large.
Since   $Var(l)$  exceed  the limiting values  $Var(\infty)$  for $2 < m  < 9$
(see Figure 2  for $m=8$)  the restriction (\ref{G1}) on variation of $G$ is not satisfied for
$m =3, ..., 8$ and we are led to unphysical results.
 This is why we consider in what follows just the cases $m= 9, 10,12,...$.

\subsection{Infinite series of solutions for $m =9$}

Now we consider the case $\bf m=9$.  We get
the following relations.

\bigskip

\begin{minipage}[h]{0.49\linewidth}
For  $\bf l=2679$:

 $\bf H\approx 0.3531582111$,

 $\bf h \approx -0.0007910955039$,

 $\bf Int \approx -0.000395588$,

 $\bf Var \approx 0.001120145$.

\medskip

 For  $\bf l=2680$:

 $\bf H \approx 0.3531583594$,

$\bf h \approx -0.0007908005933$,

$\bf Int \approx -0.000395434$,

$\bf Var \approx 0.001119706$.

\end{minipage}
\begin{minipage}[h]{0.49\linewidth}
For  $\bf l=2681$:

 $\bf H\approx 0.3531585062$,

 $\bf h \approx -0.0007905059028$,

 $\bf Int \approx -0.000395288$,

 $\bf Var \approx 0.001119293$.

 \medskip

 For  $\bf l=2682$:

 $\bf H \approx 0.3531586532$,

$\bf h \approx -0.0007902114318$,

$\bf Int \approx -0.000395141$,

$\bf Var \approx 0.001118876$.
\end{minipage}

\bigskip

We do not present here exact analytical forms of these solutions in radicals which are bulky ones. For example, the relation for the parameter $H$, when   $l=2680$, contains ($17$  times) the radical
$\sqrt{32839778319264444823234828568184005}$.

The numerical calculations for fixed $m=9$ gives us an evidence of monotonically decreasing behaviour of the function $Var(l)$ for $l \geq 2680$
 \fnm[5]\fnt[5]{A rigorous analytical proof of this fact may be  a subject of a separate work.}
  as well as the asymptotical relation:  $Var(l) \sim  A/l$, as $l \to + \infty$, where $A >0$. See Figure 3.

Thus, for $m=9$   there is an infinite series of
admissible cosmological solutions with $l=2680,2681, ...$, which  satisfy all the conditions imposed. Any of such solution describes  accelerated
expansion of the three-dimensional factor space with sufficiently small  value of the variation of the effective gravitational constant $G$. This variation may be arbitrary small for a big enough value of $l$.

\begin{figure}[h]
\begin{minipage}[h]{0.49\linewidth}
\center{\includegraphics[width=1\linewidth]{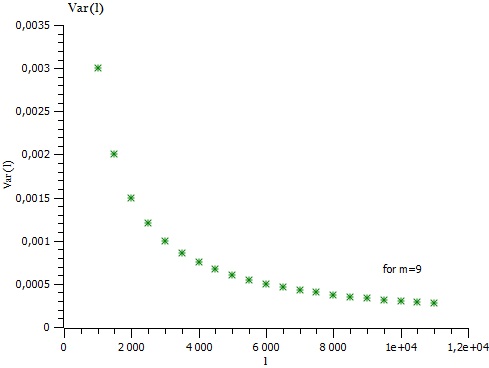}}
\end{minipage}
\hfill
\begin{minipage}[h]{0.49\linewidth}
\center{\includegraphics[width=1\linewidth]{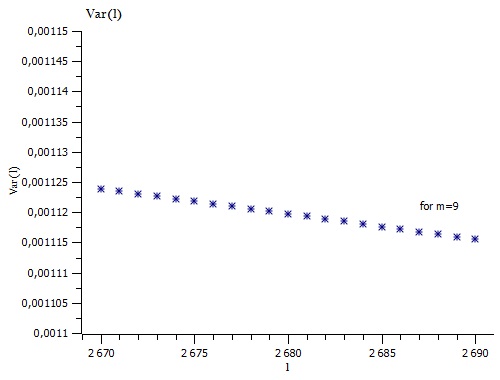}}
\end{minipage}
\caption{Behavior of the variation of $G$ parameter  $Var$ for fixed $m=9$  while  $l$ is changing from $1000$ to $11000$ by $500$ (left) and from $2670$ to $2690$ by $1$ (right).}
\label{ris:image2}
\end{figure}
\vfill

 The infinite series of solutions for $m= 9$  and  $l = 2680, 2681, ...$  starts from the (special) total  dimension  $D=2690$.
 For  $D <  2690$ and $m=9$  the solutions do not obey restriction (\ref{G1}) on variation of $G$ and hence are not of interest
 for our  consideration.

\subsection{Some  solutions for $m >9$ with  minimal $Var$-parameter   }

When $m>9$ the internal space parameter $Int$ becomes positive.
As $l$ can only be a natural number we should look for a value of $l$ which gives
the minimal magnitude of the variation of $G$ parameter $Var$.
Below we present the calculated values of our parameters
($H$, $h$, $Int=(m-3)\cdot H+l\cdot h$ and $Var=|(m-3)+\frac{l\cdot h}{H}|$) for each of  considered cases:

\begin{enumerate}
\item for {\bf $\bf m=10$}:

\begin{figure}[h]
\begin{minipage}[h]{0.49\linewidth}
{\it  the variation of $G$ parameter is minimal for {$\bf l=31$}, see Figure 4,}

\medskip
 {$\bf H \approx 0.2996055415$},

\medskip
 {$\bf h \approx -0.06764217686$},

\medskip
 ${\bf Int \approx 0.000331307}$,

\medskip
 ${\bf Var \approx 0.001105812}$.

\medskip

{\it This case is not of particular interest. The radical forms of the solutions are too bulky,
so we approximated them. Nevertheless the variation of $G$ parameter is out of the allowed domain
and the ``internal space''\\  parameter is positive.}

\end{minipage}
\begin{minipage}[h]{0.49\linewidth}
\center{\includegraphics[width=1\linewidth]{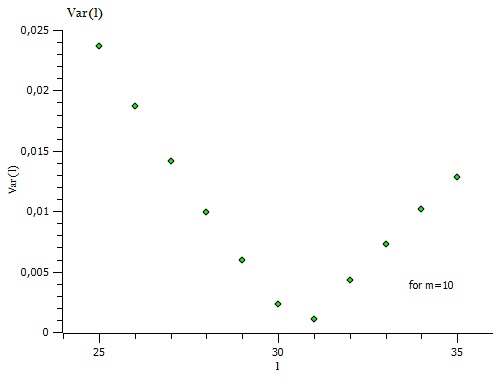}}
\caption{The variation  of $G$ parameter for $m=10$.}
\end{minipage}
\label{ris:image3}
\end{figure}
\vfill

\clearpage

\item for {\bf $\bf m=11$}:
\begin{figure}[h]
\begin{minipage}[h]{0.49\linewidth}
{\it the variation of $G$  and the ``internal space''

 parameters are zero for {$\bf l=16$}, see Figure 5,}

\medskip
 {$\bf H=\sqrt{\frac{1}{15}}$},

\medskip
 {$\bf h=-\frac{1}{2}\sqrt{\frac{1}{15}}$},

\medskip
 ${\bf Int=0}$,

\medskip
 ${\bf Var=0}$.

\medskip

{\it This case is the first one with zero variation of $G$. Also, the exact values of ``Hubble-like''
parameters ($H=-2h$) in contrast to the previous case have rather simple and \\ compact form.}

\end{minipage}
\begin{minipage}[h]{0.49\linewidth}
\center{\includegraphics[width=1\linewidth]{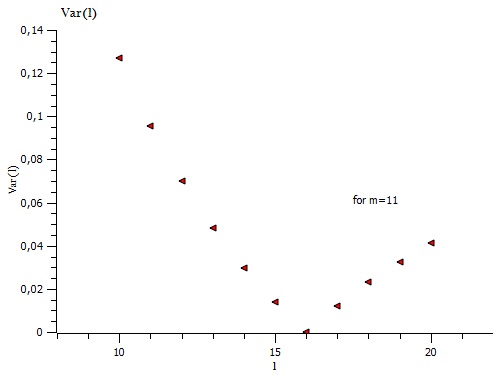}}
\caption{The variation of $G$ parameter for $m=11$.}
\end{minipage}
\label{ris:image4}
\end{figure}
\vfill

\item for {\bf $\bf m=12$}:

\begin{figure}[h]
\begin{minipage}[h]{0.49\linewidth}
{\it the variation of $G$ parameter is minimal for {$\bf l=11$}, see Figure 6,}

\medskip
 {$\bf H\approx0.2264080186$},

\medskip
 {$\bf h \approx -0.1852491999$},

\medskip
 ${\bf Int \approx -0.000069032}$,

\medskip
 ${\bf Var\approx 0.000304899}$.

\medskip

{\it Here  all our four conditions are satisfied.
The variation of $G$ parameter is non-zero and the volume of the internal space is \\ decreasing.}

\end{minipage}
\begin{minipage}[h]{0.49\linewidth}
\center{\includegraphics[width=1\linewidth]{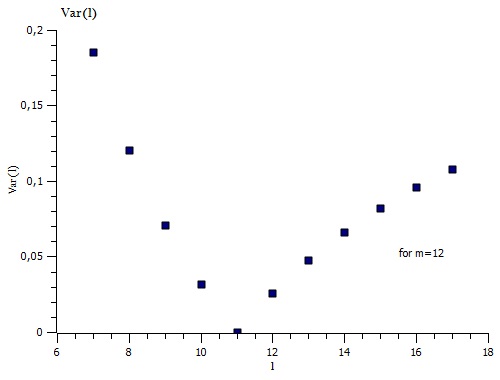}}
\caption{The variation of $G$ parameter for $m=12$.}
\end{minipage}
\label{ris:image5}
\end{figure}
\vfill
\clearpage

\item for {\bf $\bf m=13$}:

\begin{figure}[h]
\begin{minipage}[h]{0.49\linewidth}
 {\it the variation of $G$ parameter is minimal for {$\bf l=9$} and
 the ``internal space'' parameter is positive, see Figure 7,}

\medskip
 {$\bf H \approx0.2039802, h\approx-0.2261006$},

\medskip
 ${\bf Int \approx0.0048967, Var\approx0.0240058}$,

\bigskip
 {\it for {$\bf l=8$} the variation of $G$ parameter is slightly higher,
 but the ``internal space'' parameter is negative, see Figure 8, }

\medskip
 {$\bf H \approx0.1942063, h \approx-0.2498379$},

\medskip
 ${\bf Int \approx- 0.0052611, Var \approx 0.0263919}$.

{\it Both cases are excluded by $G$-dot\\ restrictions.}

\end{minipage}
\begin{minipage}[h]{0.49\linewidth}
\center{\includegraphics[width=1\linewidth]{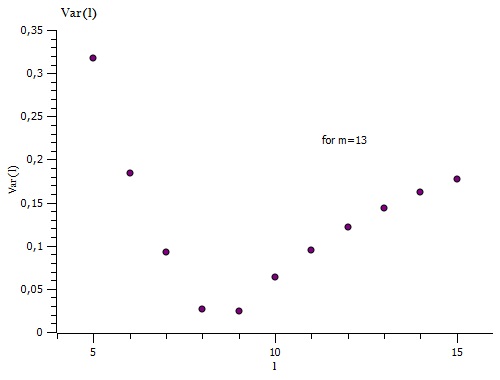}}
\caption{The variation of $G$ parameter for $m=13$.}
\end{minipage}
\label{ris:image6}
\end{figure}
\vfill

\item for {\bf $\bf m=14$}:

\begin{figure}[h]
\begin{minipage}[h]{0.49\linewidth}
{\it the variation of $G$ parameter is minimal  for {$\bf l=7$}, see Figure 8,}

\medskip
 {$\bf H\approx0.1822582965$},

\medskip
 {$\bf h\approx-0.2863787788$},

\medskip
 ${\bf Int \approx 0.000189810}$,

\medskip
 ${\bf Var\approx0.00104143}$.

\medskip

{\it The  variation of $G$ parameter  exceeds our limits,
and the condition of the volume contraction of the ``inner space'' is not met.}

\end{minipage}
\begin{minipage}[h]{0.49\linewidth}
\center{\includegraphics[width=1\linewidth]{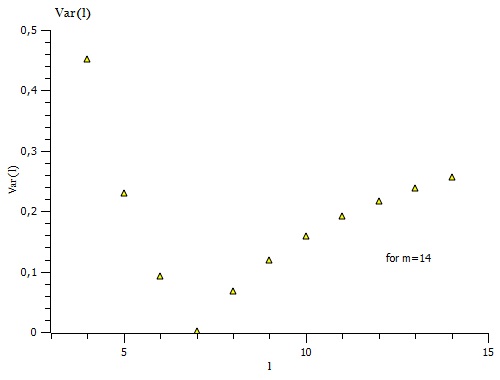}}
\caption{The variation of $G$ parameter for $m=14$.}
\end{minipage}
\label{ris:image7}
\end{figure}
\vfill
\clearpage

\item for {\bf $\bf m=15$}:
\begin{figure}[h]
\begin{minipage}[h]{0.49\linewidth}
{\it the variation of $G$ and the ``internal space''

parameters are zero for {$\bf l=6$},  see Figure 9, }

\medskip
 {$\bf H=\frac{1}{6}$},

\medskip
 {$\bf h=-\frac{1}{3}$},

\medskip
 ${\bf Int=0}$,

\medskip
 ${\bf Var=0}$.

\medskip

{\it This is the second case with zero variation of $G$.
The exact values of ``Hubble-like'' parameters ($H=-\frac{1}{2}h$) have simple and compact form.}

\end{minipage}
\begin{minipage}[h]{0.49\linewidth}
\center{\includegraphics[width=1\linewidth]{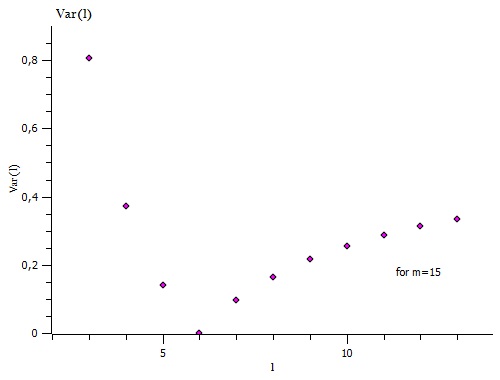}}
\caption{The variation of $G$ parameter for $m=15$.}
\end{minipage}
\label{ris:image8}
\end{figure}

\end{enumerate}
\vfill

Now we will reverse our method and look for the solutions with minimal variation of $G$ for fixed $l$ instead of $m$. The calculations lead to the following rule:
the lesser is $l$ the greater is an appropriate $m$ which gives the  minimum of $Var$ .
As we consider $l \ge 3$ and for $l=6$ the solution with minimal variation of $G$ parameter
is already found, we should examine only three cases:

\begin{enumerate}

\item for {$\bf l=5$}:

\begin{figure}[h]
\begin{minipage}[h]{0.49\linewidth}
{\it the variation of $G$ parameter is minimal for {$\bf m=17$}, see Figure 10,}

\medskip
 {$\bf H \approx 0.1447364880$},

\medskip
 {$\bf h \approx -0.4041874693$},

\medskip
 ${\bf Int \approx 0.005373486}$,

\medskip
 ${\bf Var \approx 0.017555134}$.

\medskip

{\it None of the desirable conditions are satisfied.}

\end{minipage}
\begin{minipage}[h]{0.49\linewidth}
\center{\includegraphics[width=1\linewidth]{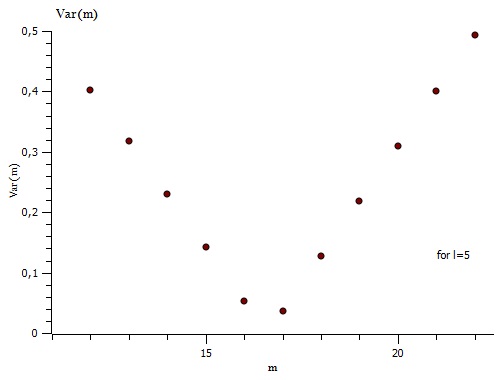}}
\caption{The variation of $G$ parameter for $l=5$.}
\end{minipage}
\label{ris:image9}
\end{figure}
\vfill
\clearpage

\item for {\bf $\bf l=4:$}

\begin{figure}[h]
\begin{minipage}[h]{0.49\linewidth}
- {\it the variation of G parameter is minimal for {$\bf m=20$}, see Figure 11,}

\medskip
- {$\bf H\approx0.1220672556$}

\medskip
- {$\bf h\approx-0.5176845111$}

\medskip
- ${\bf Int\approx0.004405301}$

\medskip
- ${\bf Var\approx0.03608913}$

\medskip

{\it The amount of variation is too high and the ``internal space'' parameter is positive.}

\end{minipage}
\begin{minipage}[h]{0.49\linewidth}
\center{\includegraphics[width=1\linewidth]{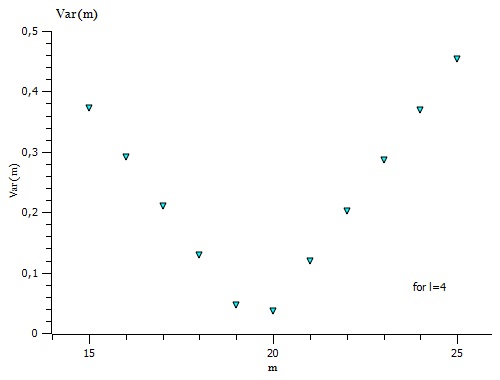}}
\caption{Variation of $G$ for $l=4$.}
\end{minipage}
\label{ris:image10}
\end{figure}
\vfill

\item for {\bf $\bf l=3$}:

\begin{figure}[h]
\begin{minipage}[h]{0.49\linewidth}
 {\it  the variation of $G$ parameter is minimal for {$\bf m=28$}, see Figure 12,}

\medskip
 {$\bf H \approx 0.09202826388$},

\medskip
 {$\bf h \approx -0.7765606872$},

\medskip
 ${\bf Int \approx 0.000272078}$,

\medskip
 ${\bf Var \approx 0.00295646}$.

\medskip

{\it The ``internal space'' parameter is also positive and the variation of $G$ parameter  exceeds the limits imposed.}

\end{minipage}
\begin{minipage}[h]{0.49\linewidth}
\center{\includegraphics[width=1\linewidth]{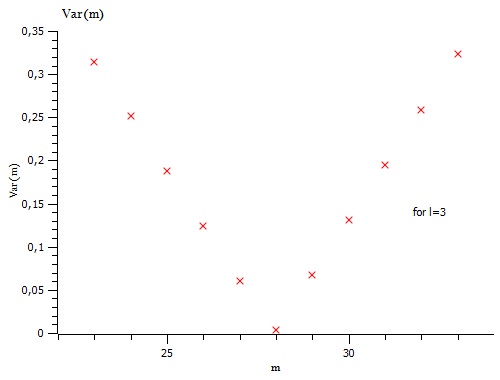}}
\caption{The variation of $G$ parameter for $l=3$.}
\end{minipage}
\label{ris:image11}
\end{figure}
\end{enumerate}

 Thus, in this subsection we have obtained  cosmological solutions for $m > 9$,
 which satisfy all four conditions for the following cases:
  \begin{enumerate}
  \item {$\bf m=11,l=16$} (zero variation of $G$);
  \item {$\bf m=12,l=11$};
  \item {$\bf m=15,l=6$} (zero variation of $G$).
  \end{enumerate}

It should be noted that for $m =3$ and  $l = 2$ the  solution with $H \approx 0,750173$ and $h \approx - 0,541715$ was found earlier in  \cite{Rat}. For this solution we have a contracting ``internal space'' but  the variation of $G$ is a huge one ($\dot{G}/G$ is of Hubble parameter order). Recently, an exact analytic form of this solution was obtained in \cite{ChPavTop}.

 \section{Conclusions}

 We have considered the  $(n +1)$-dimensional  Einstein-Gauss-Bonnet (EGB) model.
 By using the  ansatz with diagonal  cosmological type metrics,
 we have found solutions with exponential dependence of scale factors
  with respect to ``synchronous-like'' variable $\tau$.

  In cosmological case ($w= -1$) these
  solutions   describe   an exponential expansion of ``our'' 3-dimensional factor-space with
  the Hubble parameter $H > 0$   and obey  the observational constraints on the temporal variation of the effective gravitational constant $G$. Any solution describes $(m-3 + l)$-dimensional
  ``internal space'', which is anisotropic: it is expanding in $(m-3)$ dimensions with the Hubble rate
$H > 0$ and contracting  in $l$ dimensions.

  These solutions were found (in numerical or analytical forms) for the following cases:
  \begin{enumerate}
   \item {$\bf m=9, l \ge 2680$} (variation of $G$ tends to $0$ as $l \to + \infty$);
   \item {$\bf m=11,l=16$} (variation of $G$ is zero);
  \item {$\bf m=12,l=11$};
  \item {$\bf m=15,l=6$} (variation of $G$ is zero).
   \end{enumerate}

Thus, we have shown that  it is possible in the framework of EGB model to describe the accelerated
expansion of the three-dimensional factor space with sufficiently small (or even zero) value of
the variation of the effective gravitational constant $G$.
For the case $w = 1$ we have obtained  by product a family of static configurations which
may be of interset within some other possible applications.

Here we have  considered a gravitational model in more than 4-dimensions.
In such a case the Gauss-Bonnet term gives non-trivial contributions to the
generalized Einstein field equations. In particular we have shown that there are
cosmological solutions in agreement with observations when ``projected'' on the
$(3+1)$-dimensional physical space-time. For the sake of simplicity,  we restrict ourselves to
vacuum solutions in multidimensional gravity with the Gauss-Bonnet term .
Such ansatz  may be considered as a part of  a general ``geometrical program''
aimed at the explanation of dark energy in 4-dimensional space,
e.g. by using extra dimensions  and modified equation of motions
just without matter sources. This is a first step. The inclusion of matter sources
(e.g. anisotropic fluid) will be the next step,  e.g.  as a subject of a next publication.

\end{document}